\documentclass[11pt,a4paper]{article}

\usepackage{amsmath,amssymb}
\usepackage{epsfig,graphicx}
\usepackage{subfigure}
\usepackage{graphicx}
\usepackage{rotating}
\usepackage{cancel}
\usepackage{bm}
\usepackage{color}
\usepackage{comment}
\usepackage{cite}
\usepackage{psfrag}

\renewcommand\[{\left[}

\newcommand{\exclude}[1]{}
\newcommand{\xalp}{x_{\rm ALP}}
\newcommand{\falp}{f_{\rm ALP}}
\newcommand{\malp}{m_{\rm ALP}}

\def\beq{\begin{equation}}
\def\eeq{\end{equation}}

\begin{document}
\numberwithin{equation}{section}
\title{{\normalsize  \mbox{}\hfill DESY 14-023, MPP-2014-41}\\
\vspace{2.5cm} 
\Large{\textbf{A 3.55 keV hint for decaying axion-like particle dark matter
\vspace{0.5cm}}}}

\author{Joerg Jaeckel$^{1}$, Javier Redondo$^{2,3}$ and Andreas Ringwald$^{4}$\\[2ex]
\small{\em $^1$Institut f\"ur theoretische Physik, Universit\"at Heidelberg, Philosophenweg 16, 69120 Heidelberg, Germany}\\[0.5ex]
\small{\em $^2$Arnold Sommerfeld Center, Ludwig-Maximilians-Universit\"at,  80333 M\"unchen, Germany}\\[0.5ex]
\small{\em $^3$Max-Planck-Institut f\"ur Physik, M\"unchen, 80805 Germany}\\[0.5ex] 
\small{\em $^4$Deutsches Elektronen Synchrotron DESY, Notkestrasse 85, 22607 Hamburg, Germany}\\[0.5ex]
}

\date{}
\maketitle

\begin{abstract}
\noindent
Recently, indications for an emission line at $3.55\,{\rm keV}$ have been found in the combined spectra 
of a large number of galaxy clusters and also in Andromeda.
This line could not be identified with any known spectral line. It is tempting to speculate that it has its origin in the decay of a particle contributing all or part of the dark matter. In this note we want to point out that axion-like particles being all or part of the dark matter are an ideal candidate to produce such a feature. More importantly the parameter values necessary are quite feasible in extensions of the Standard Model based on string theory and could be linked up to a variety of other intriguing phenomena, which also potentially allow for new tests of this speculation.  
\end{abstract}

\newpage

\section{Introduction}
Two groups~\cite{Bulbul:2014sua,Boyarsky:2014jta} have recently found a new emission line in the spectra of galaxy clusters and also in a single galaxy, Andromeda~\cite{Boyarsky:2014jta}.
This line is at an energy of $3.55\,\,{\rm keV}$ and so far has not been identified with a transition in any known element.
Although the statistical significance is at the $(4-5)\sigma$-level, the line is very weak and a number of systematic effects could 
affect the interpretation. 
Keeping this caveat in mind, it is nevertheless tempting to speculate that this line could be a sign of new physics, more precisely the sign of a decaying form of dark matter. 
Indeed, the authors of~\cite{Boyarsky:2014jta} have found indications that the line becomes stronger towards the center of the cluster, which is consistent with the idea of a decaying dark matter particle.

Both~\cite{Bulbul:2014sua,Boyarsky:2014jta} as well as~\cite{Ishida:2014dlp} provide a possible explanation in the form a sterile neutrino decaying into a photon and an active neutrino. Another possibility could be eXciting Dark Matter~\cite{Finkbeiner:2014sja}.
In this brief note we point out, similar to~\cite{Higaki:2014zua}, that the decay of an axion-like particle (ALP) being all or part of the dark matter is another simple explanation. 
The strength of the coupling to two photons required can be easily motivated in extensions of the Standard Model based on string theory.

String theory often also features a number of additional ALPs, which generically have similar couplings. Thus, one can link the 3.55 keV line to a variety of other puzzling observations. While this may increase the level of speculation it provides additional handles to test the hypothesis.

\section{A simple ALP explanation}\label{simple}
The defining feature of axion-like particles~\cite{Masso:1995tw,Masso:1997ru,Masso:2004cv,Jaeckel:2006xm} is their coupling to two photons,
\begin{equation}
{\mathcal L}_{\rm int}=\frac{1}{4}g_{\phi \gamma\gamma}\phi F^{\mu\nu}\tilde{F}_{\mu\nu},
\end{equation}
where for concreteness we have taken the ALP to be a pseudoscalar (the scalar case is analogous).

This interaction allows the ALP to decay into two photons with a lifetime of
\begin{equation}
\tau_{\phi}=\frac{1}{\Gamma_{\gamma\gamma}}=\frac{64\pi}{g^{2}_{\phi\gamma\gamma} m^{3}_{\phi}}.
\end{equation}

Assuming that the ALP makes up all of the dark matter~\cite{Arias:2012az} the photon fluxes found in~\cite{Bulbul:2014sua,Boyarsky:2014jta} correspond to lifetimes in the range\footnote{Note that there is a difference of a factor of two in the lifetime compared to the sterile neutrino interpretation because the ALP decays into two photons.}
\begin{equation}
\tau_{\phi}\sim (4\times 10^{27}-4\times 10^{28}) \,{\rm s} . 
\end{equation}

\begin{figure}[t]
    \begin{center}
     \includegraphics[scale=0.5]{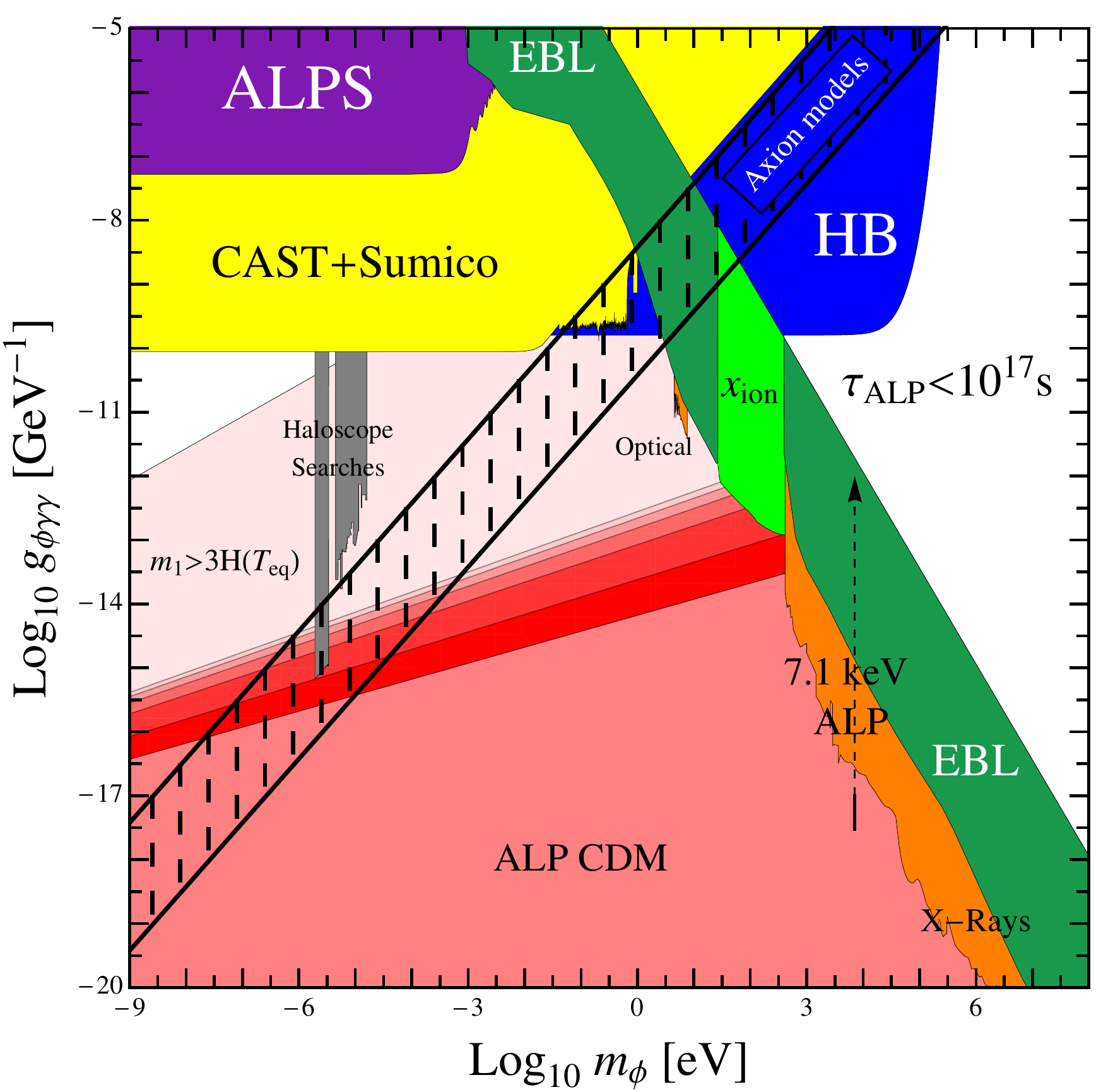}
        \caption{Parameter space for axion-like particle dark matter~\cite{Arias:2012az}. The thick black line indicates the region which could explain the 3.55 keV line assuming that the 7.1 keV ALP makes up all of the dark matter.
         The dashed black line shows the possible couplings when the ALP is only a fraction of the dark matter. The limits shaded in green and in orange assume that the ALP gives the full dark matter density. Accordingly, they weaken with the dark matter fraction $\sim 1/\sqrt{\xalp}$.}
\label{recyclefig}
    \end{center}
\end{figure}

Combining this with the energy of the photon line, we find the following parameters,
\begin{equation}
\malp\sim 7.1\,{\rm keV},\qquad g_{\phi\gamma\gamma}\sim (3-10)\times 10^{-18}\,{\rm GeV}^{-1}.
\end{equation}
These parameters are not excluded by any of the existing constraints~\cite{Cadamuro:2011fd,Arias:2012az}. In fact ALPs with these parameter values can be produced in the early Universe via the misalignment mechanism and can be the cold dark matter~\cite{Arias:2012az}.
The suitable parameter range together with existing constraints is shown in Fig.~\ref{recyclefig} as the vertical black line.

Alternatively we could think that ALPs with these parameter values make up only a fraction of the total dark matter energy density,
\begin{equation}
\xalp=\frac{\rho_{\rm ALP}}{\rho_{\rm DM}}.
\end{equation}
In this case, the required lifetime (as long as it is larger than the age of the Universe) decreases as
\begin{equation}
\tau_{\phi}\sim \xalp
\end{equation}
and the required coupling increases as
\begin{equation}
g_{\phi\gamma\gamma}\sim \frac{1}{\xalp^{1/2}}.
\end{equation}

Requiring that the lifetime exceeds that of the 
Universe\footnote{One can possibly use somewhat smaller lifetimes, but already with 
the lifetime of the Universe one should see systematically larger signals in observations at high red-shift.}, one can increase the coupling in this way up to (see also Fig.~\ref{recyclefig})
\begin{equation}
g^{\rm max}_{\phi\gamma\gamma}\sim 10^{-12}\,{\rm GeV}, \qquad \xalp^{\rm max}\sim 10^{-10}.
\end{equation}
One may think that this corresponds to an additional fine-tuning. However this is not the case for ALPs 
produced via the misalignment mechanism.
Actually, ALP CDM already requires a tuning of the initial angle
\begin{equation}
\theta_{\rm ini}\sim \frac{\phi_{\rm ini}}{\falp}
\end{equation}
such that the produced dark matter density (for the full formula see Eq.~\eqref{darkmatterdensity})
\begin{equation}
\Omega_{\rm ALP}\sim \theta^{2}_{\rm ini} \falp^2
\end{equation}
gives the desired value. Here $\falp$ is the ALP decay constant, for instance specifying the scale at which the ALP is generated as a pseudo-Goldstone boson. 

At a given level of tuning for $\theta_{\rm ini}$, the DM density decreases with decreasing $\falp$.
Now we just need to note that the coupling and $\falp$ are related via,
\begin{equation}
g_{\phi\gamma\gamma}\sim \frac{\alpha}{2\pi\falp}, 
\end{equation}
and the flux is
\begin{equation}
{\rm flux}\sim g^{2}_{\phi\gamma\gamma}\Omega_{\rm ALP}\sim \theta^{2}_{\rm ini}, 
\end{equation}
independent of the size of $\falp$ and therefore the coupling.

Making use of this freedom, the ALP decay constant -- and correspondingly the couplings -- can be in the range,
\begin{equation}
\falp\sim (10^{9}-4\times10^{14})\,{\rm GeV},\qquad g_{\phi\gamma\gamma}\sim (3\times 10^{-18}-10^{-12})\,{\rm GeV}^{-1},
\end{equation}
without any increase in finetuning compared to the case where the ALP makes up all of the dark matter.
The region where the ALP is a sub-dominant component is shown as the vertical dashed black line in Fig.~\ref{recyclefig}.

In general, axion-like particles could lead to observable isocurvature perturbations in the cosmic microwave background. This could be a cosmological signal for this interpretation
of the 3.55 keV line. Beyond that one could have signals in the form of non-Gaussianities generated by isocurvature fluctuations. In the case where 
the ALP is sub-dominant the non-Gaussianity could be visible even before one detects the isocurvature modes themselves~\cite{Kawasaki:2008sn}. However, note that the isocurvature perturbations are suppressed in some particular models~\cite{Folkerts:2013tua}.

Interestingly, the emission of ALPs with a decay constant at the upper end of this range from Red Giant (RG) stars, could be responsible for the small amount of non-standard stellar cooling that seems to be present in the RG population of the globular cluster M5~\cite{Viaux:2013lha}, if we use 
\begin{equation}
g_{aee}\sim \frac{m_{e}}{\falp}
\end{equation}
and compare to the upper limit of $g_{aee}=4.3\times 10^{-13}$.
However, the extra cooling observed in white dwarfs~\cite{Isern:2012ef,Corsico:2012ki,Isern:2013lwa} cannot be explained with an ALP of this  mass, since the available temperatures in a white dwarf are too low for effective production.

\section{Thermally produced ALPs}
In the simplest models and cosmologies, the misalignment mechanism is extremely efficient to produce ALP DM. 
Actually, it tends to overproduce it in the parameter range motivated by the 3.55 keV line. 
However, one can imagine cases in which this contribution can be strongly suppressed.  

For instance, one could have initially very large masses arising from thermal effects, couplings to additional moduli fields or non-minimal gravitational couplings. The energy density in the ALP field gets diluted by the expansion (behaving as dark matter) only after the Hubble rate becomes comparable to the mass, $H\sim m_\phi$, and thus high masses in the early universe trigger the dilution earlier than the bare mass today, with the consequent decrease in abundance.  

Another example happens for the largest values of $g_{\phi\gamma\gamma}$, where primordial magnetic fields can trigger the mixing of the ALP condensate into a DC electric field which can be efficiently discharged by the huge conductivity of the primordial plasma~\cite{Ahonen:1995ky}. 

However, the relic abundance of ALPs can also be thermally produced. 
The coupling to two photons provides an ALP production channel by the Primakoff process $q+\gamma\to q+ \phi$ where $q$ is any charged particle in the plasma. This 
generates ALPs at a rate \cite{Bolz:2000fu,Cadamuro:2011fd}
\begin{equation}
\Gamma_Q=\frac{\alpha g^2_{\phi\gamma\gamma}\pi^2 }{36\zeta(3)}\left(\log\left(\frac{T^2}{m_\gamma^2}\right)+0.82\right) n_q, 
\end{equation}
where $n_q$ is the	effective number density of charged	particles,  $n_q	=	\sum_i Q^2_i n_i \equiv 
(\zeta(3)/\pi^2)g_q(T)T^3$, $Q_i$ is the charge of i-th particle species, and the parameter $g_q(T)$
represents the effective number of relativistic {\em charged} degrees of freedom. 
At very high temperatures, we shall employ the ALP couplings to electroweak bosons $W$ and $B$ as in \cite{Salvio:2013iaa} 
but already the photon coupling gives us an idea of the phenomenological consequences.  
In particular, the rate is proportional to $T^3$ and thus redshifts faster than the Hubble expansion rate $H\sim T^2/m_{\rm Pl}$. 
Therefore, the highest temperatures of the early universe determine whether a full thermal population is established. 
The contribution of a fully thermalised species to the DM abundance is \cite{Cadamuro:2011fd}
\begin{equation}
\frac{\rho_{\rm ALP}}{\rho_{\rm  CDM}}=\frac{m_\phi}{154\rm eV}\frac{106.75}{g_\star(T_f)}, 
\end{equation}
where $T_f$ is the temperature at which $\Gamma/H$ becomes smaller than 1 and the ALP interactions freeze out. 
Since $m_\phi\sim 7.1$ keV, we should never reach thermal equilibrium or we would have too much DM --- unless of course 
$g_\star$ is humongous. 
Therefore, the reheating temperature of the universe, $T_R$, should be below the freeze out temperature, $T_f$. 
The relic abundance in this ``freeze in'' scenario\footnote{The ``freeze in" production of feebly interacting massive partices (FIMPs) has been considered recently by a number of authors in different models~\cite{Hall:2009bx,Mambrini:2013iaa,Blennow:2013jba}. } is suppressed with respect to the thermal case by a factor $\sim \left.\Gamma/H\right|_{T_R}$
\begin{equation}
\frac{\rho_{\rm ALP}}{\rho_{\rm  CDM}}\sim \frac{m_\phi}{154\rm eV}\frac{106.75}{g_\star(T_R)}\left.\frac{\Gamma}{H}\right|_{T_R}. 
\end{equation}
This equation relates the DM fraction, the coupling  $g_{\phi\gamma\gamma} $ (implicit in $\Gamma$) and the reheating temperature. 
As discussed in section \ref{simple}, the requirement that the ALP decays account for the whole 3.55 keV flux fixes a relation between the ALP DM fraction and the coupling, $g_{\phi\gamma\gamma}$,  
\begin{equation}
x_{\rm ALP} \left(\beta\frac{10^{-17}\rm GeV^{-1}}{g_{\phi\gamma\gamma}}\right)^2= 1, 
\end{equation}
where $\beta\sim  0.3-1$. Therefore we can predict the required reheating temperature of the universe as a function of $g_{\phi\gamma\gamma}$ (or $x_{\rm ALP}$)
\begin{equation}
T_R\sim 10^{16}\, {\rm GeV} \left(\frac{10^{-17}\rm GeV^{-1}}{g_{\phi\gamma\gamma}}\right)^4\beta^2\left(\frac{g_\star(T_R)}{106.75}\right)^{3/2} 
\frac{30}{g_q(T_R)} . 
\end{equation}
Values at the higher end of the coupling range require very low reheating temperatures. Indeed the smallest values we have considered $\sim 10^{-12}$ GeV$^{-1}$ are in conflict with the absolute lower limits on the reheating temperature $\sim 4$ MeV~\cite{Hannestad:2004px} but the dependence is so steep that already $10^{-13}$ GeV$^{-1}$ is allowed. 

We note, however, that for very high reheating scales one also needs a fairly high scale of inflation which again can lead to observable or, even too large
isocurvature fluctuations. This happens roughly at a reheating temperature of order $10^{11}\,{\rm GeV}$.
So thermal production in this sense works easiest for moderately large $g_{\phi\gamma\gamma}$ or
equivalently moderately small $\falp$.  

\section{ALP + hidden photon requires no finetuning}
As we have already mentioned, some amount of finetuning is needed in order to achieve the correct density of the 7.1 keV ALPs.
The amount of ALP dark matter produced via the misalignment mechanism (and whose mass is not influenced by the thermal environment) is given by,
\begin{eqnarray}
\label{darkmatterdensity}
\frac{\rho_{\rm ALP}}{\rho_{\rm CDM}}&=&0.4\times \left(\frac{{\mathcal{F}}(T_{1})}{0.5}\right) \left(\frac{\malp}{7.1\,{\rm keV}}\right)^{\frac{1}{2}}
\left(\frac{\falp}{10^{10}\,{\rm GeV}}\right)^2\theta^{2}_{\rm ini},
\end{eqnarray}
where ${\mathcal{F}}(T_{1})=(g_{\star}(T_{1})/3.36)^{3/4}/(g_{\star,S}(T_{1})/3.91)$ with $g_{\star}$ and $g_{\star,S}$ the number of energy and entropy degrees of freedom evaluated at $H(T_1)=m_\phi$, respectively~\cite{Arias:2012az}.

Therefore, for a decay constant $\falp\sim (10^{9}-10^{11})\,{\rm GeV}$, the ALP density is of the same order of magnitude as the observed dark matter density, without needing to choose unnaturally small values of the initial angle $\theta_{\rm ini}$.
However, as discussed above this naively corresponds to a coupling to photons of the order of
\begin{equation}
g_{a\gamma\gamma}\sim\frac{\alpha}{2\pi \falp}\sim (10^{-14}-10^{-12})\,{\rm GeV}^{-1},
\end{equation}
which, if the ALPs make up all of the DM, is way too large to re-produce the observed 3.55~keV line.

A simple way to remedy this problem is if the ALP resides in a hidden sector and directly only (or at least dominantly) couples to an extra U(1) gauge boson (as in~\cite{Masso:2005ym}), $X^{\mu}$ with field strength $X^{\mu\nu}$. In this case we have,
\begin{equation}
{\mathcal{L}}^{\rm HS}_{\rm int}=\frac{1}{4}g_{\phi XX}\phi X^{\mu\nu}\tilde{X}_{\mu\nu},
\end{equation}
with
\begin{equation}
g_{\phi XX}\sim \frac{\alpha_{X}}{2\pi \falp},
\end{equation}
where $\alpha_{X}$ is the hidden U(1) equivalent of the fine-structure constant, $e_X^2/4\pi$, with $e_X$ the hidden U(1) gauge coupling. 

It is now quite natural~\cite{Dienes:1996zr,Lukas:1999nh,Abel:2003ue,Blumenhagen:2005ga,Abel:2006qt,Abel:2008ai,Goodsell:2009pi,Goodsell:2009xc,Goodsell:2010ie,Heckman:2010fh,Bullimore:2010aj,Cicoli:2011yh,Goodsell:2011wn} that the hidden U(1) mixes with the photon via a so-called kinetic mixing~\cite{Holdom:1985ag},
\begin{equation}
{\mathcal{L}}^{\rm kin\,mix}=-\frac{1}{2}\chi F^{\mu\nu}X_{\mu\nu}.
\end{equation}
After removing this mixing by a field re-definition, 
\begin{equation}
X^{\mu}\rightarrow X^{\mu}+\chi A^{\mu}, 
\end{equation}
we have the following interactions of our ALP with the ordinary photon,
\begin{equation}
{\mathcal{L}}^{\rm SM}_{int}=\frac{1}{2}g_{\phi \gamma X}\phi F^{\mu\nu}\tilde{X}_{\mu\nu}+\frac{1}{4}g_{\phi \gamma \gamma}\phi F^{\mu\nu}\tilde{F}_{\mu\nu},
\end{equation}
where
\begin{equation}
g_{a\gamma X}= \chi g_{a XX}\sim \chi \frac{\alpha_{X}}{2\pi \falp},\qquad g_{a\gamma \gamma}= \chi^2 g_{a XX}\sim \chi^2 \frac{\alpha_{X}}{2\pi \falp}.
\end{equation} 

For small $\chi$ the dominant decay is that into two $X$, but we also have a decay $\phi\to \gamma+X$ which can give us the photon line (the decay to two $\gamma$ is further suppressed by the small $\chi$).

Choosing $\alpha_{X}\sim (0.1-10)\alpha $ values in the 
\begin{equation}
\chi\sim 10^{-7}-10^{-2}
\end{equation}
can now reproduce the 3.55 keV line while requiring no finetuning of the ALP density.

The extra U(1), ``hidden photon'' can be massless or massive.
The massless case is perhaps simplest and is not constrained by any limits known to us. In the massive case,  a large part of the parameter space is already excluded for masses up to the keV region (see, e.g.~\cite{Jaeckel:2013ija}). Nevertheless some parts are still free and could be explored in near future experiments and observations (see~\cite{Jaeckel:2013ija}), making this a phenomenologically interesting prospect.

\section{ALPs from string theory}
Compactifications of string theory typically predict a multitude of axion-like fields. For example,  one gets an axion for each
closed sub-manifold in the extra dimensions. This can be a very large number, possibly more than a 100, a fact that has triggered the discussion about a string axiverse~\cite{Arvanitaki:2009fg,Acharya:2010zx,Cicoli:2012sz}. In these constructions one finds, next to the axion, additional axion-like particles
with masses distributed uniformly in the logarithm. This is makes it quite likely that one of them is in the keV range.

In general one can say that the decay constants are naturally either of the string scale $\falp\sim M_{\rm s}$ (if the cycles are small) or of the Planck 
scale (if they extend into the whole volume). The latter are obviously very weakly coupled and  therefore contribute only to gravitational signals.
Importantly, all the remaining ones essentially share the same decay constant.

In the LARGE volume scenario~\cite{Balasubramanian:2005zx}, an appealing option is to have intermediate string scales of the order of
\begin{equation}
M_{\rm s}\sim (10^{9}-10^{12})\,{\rm GeV}.
\end{equation}
This string scale naturally gives rise to gravity mediated supersymmetry breaking at the TeV scale.

Alternatively, some GUT-like models prefer a string scale that is higher~\cite{Blumenhagen:2009gk,Cicoli:2011qg},
\begin{equation}
M_{\rm s}\sim  10^{14}\,{\rm GeV}.
\end{equation}

\section{Linking with other phenomena}
In the previous section we have not only seen that an ALP with suitable parameter values to explain the observed spectral line can be found in
string theory, but also that it is quite naturally accompanied by additional ALPs and perhaps also the QCD axion.
Crucially these ALPs typically have a decay constant of a similar size and therefore similar coupling strength.

\subsection*{Low decay constant $\falp\sim 10^{9}\,{\rm GeV}$}
At the upper end of the allowed range of couplings we could have an ALP at $\malp\sim 7.1\,{\rm keV}$ accompanied by an additional ALP with a much smaller mass $m_{\rm ALP\,2}\lesssim 10^{-9}\,{\rm eV}$.
While the former can give rise to the 3.55 keV line the latter could explain both the additional cooling in white dwarfs~\cite{Isern:2012ef,Corsico:2012ki,Isern:2013lwa} (via a coupling to electrons of strength $\sim m_{e}/\falp$) and the anomalous transparency of the Universe to TeV gamma rays~\cite{DeAngelis:2007dy,Meyer:2013pny} (via its coupling to photons).

Moreover the second ALP, which could have slightly (by an ${\mathcal O}(1)$ factor) larger couplings, could be within the reach of near future experiments such as IAXO~\cite{Armengaud:2014gea} or ALPS-II~\cite{Bahre:2013ywa}.

This scale is also very attractive from the point of view of intermediate scale string models.

\subsection*{Medium decay  constant $\falp\sim 10^{12}\,{\rm GeV}$}
In this case, the ALP producing the 3.55 keV line could be linked to a true QCD axion, which then is in a range suitable for it to be essentially all of the dark matter in the Universe. (The fraction of DM contributed by the ALP is still very small in this range.)

An axion with a decay constant in this range could be searched with the haloscope technique as realized in ADMX~\cite{Asztalos:2003px}.

\subsection*{High decay  constant $\falp\sim 10^{14}\,{\rm GeV}$}
In this case the ALP responsible for the 3.55 keV line would contribute a sizeable fraction or all of the dark matter.
From this point of view there is no need for an axion or an additional ALP.

Nevertheless an axion with this decay constant would still solve the strong CP problem of QCD. At the same time it is likely that it would contribute an appreciable fraction of the dark matter (it actually requires some amount of tuning for it not to be too much dark matter). It could then be searched for
in an experiment based on LC circuits as suggested in~\cite{Sikivie:2013laa} or in experiments searching for a precession of nuclear spins~\cite{Budker:2013hfa}.

\section{Conclusions}
An axion-like particle (ALP) making up all or part of dark matter provides a simple explanation for the recently observed 3.55 keV line in the spectra of stacked galaxy clusters. 
The required values for the mass and coupling can be obtained in models of string theory. The latter also provide for the possibility that there also is an axion and/or additional ALPs with roughly the same coupling to photons. Using this one can find links to other puzzling astrophysical phenomena
and perhaps more importantly also to interesting experimental probes of this hypothesis.

\bibliographystyle{h-physrev5}
\bibliography{masterbib-new}

\begin{thebibliography}{10}

\bibitem{Bulbul:2014sua}
E.~Bulbul {\em et~al.},
\newblock (2014), arXiv:1402.2301.

\bibitem{Boyarsky:2014jta}
A.~Boyarsky, O.~Ruchayskiy, D.~Iakubovskyi, and J.~Franse,
\newblock (2014), arXiv:1402.4119.

\bibitem{Ishida:2014dlp}
H.~Ishida, K.~S. Jeong, and F.~Takahashi,
\newblock (2014), arXiv:1402.5837.

\bibitem{Finkbeiner:2014sja}
D.~P. Finkbeiner and N.~Weiner,
\newblock (2014), arXiv:1402.6671.

\bibitem{Higaki:2014zua}
T.~Higaki, K.~S. Jeong, and F.~Takahashi,
\newblock (2014), arXiv:1402.6965.

\bibitem{Masso:1995tw}
E.~Masso and R.~Toldra,
\newblock Phys.Rev. {\bf D52}, 1755 (1995), arXiv:hep-ph/9503293.

\bibitem{Masso:1997ru}
E.~Masso and R.~Toldra,
\newblock Phys.Rev. {\bf D55}, 7967 (1997), arXiv:hep-ph/9702275.

\bibitem{Masso:2004cv}
E.~Masso, F.~Rota, and G.~Zsembinszki,
\newblock Phys.Rev. {\bf D70}, 115009 (2004), arXiv:hep-ph/0404289.

\bibitem{Jaeckel:2006xm}
J.~Jaeckel, E.~Masso, J.~Redondo, A.~Ringwald, and F.~Takahashi,
\newblock Phys.Rev. {\bf D75}, 013004 (2007), arXiv:hep-ph/0610203.

\bibitem{Arias:2012az}
P.~Arias {\em et~al.},
\newblock JCAP {\bf 1206}, 013 (2012), arXiv:1201.5902.

\bibitem{Cadamuro:2011fd}
D.~Cadamuro and J.~Redondo,
\newblock JCAP {\bf 1202}, 032 (2012), arXiv:1110.2895.

\bibitem{Kawasaki:2008sn}
M.~Kawasaki, K.~Nakayama, T.~Sekiguchi, T.~Suyama, and F.~Takahashi,
\newblock JCAP {\bf 0811}, 019 (2008), arXiv:0808.0009.

\bibitem{Folkerts:2013tua}
S.~Folkerts, C.~Germani, and J.~Redondo,
\newblock Phys.Lett. {\bf B728}, 532 (2014), arXiv:1304.7270.

\bibitem{Viaux:2013lha}
N.~Viaux {\em et~al.},
\newblock (2013), arXiv:1311.1669.

\bibitem{Isern:2012ef}
J.~Isern {\em et~al.},
\newblock p. 158 (2011), arXiv:1204.3565.

\bibitem{Corsico:2012ki}
A.~H. Corsico {\em et~al.},
\newblock (2012), arXiv:1205.6180.

\bibitem{Isern:2013lwa}
J.~Isern, S.~Catalan, E.~Garcia-Berro, M.~Salaris, and S.~Torres,
\newblock (2013), arXiv:1304.7652.

\bibitem{Ahonen:1995ky}
J.~Ahonen, K.~Enqvist, and G.~Raffelt,
\newblock Phys.Lett. {\bf B366}, 224 (1996), arXiv:hep-ph/9510211.

\bibitem{Bolz:2000fu}
M.~Bolz, A.~Brandenburg, and W.~Buchmuller,
\newblock Nucl.Phys. {\bf B606}, 518 (2001), arXiv:hep-ph/0012052.

\bibitem{Salvio:2013iaa}
A.~Salvio, A.~Strumia, and W.~Xue,
\newblock JCAP {\bf 1401}, 011 (2014), arXiv:1310.6982.

\bibitem{Hall:2009bx}
L.~J. Hall, K.~Jedamzik, J.~March-Russell, and S.~M. West,
\newblock JHEP {\bf 1003}, 080 (2010), arXiv:0911.1120.

\bibitem{Mambrini:2013iaa}
Y.~Mambrini, K.~A. Olive, J.~Quevillon, and B.~Zaldivar,
\newblock Phys.Rev.Lett. {\bf 110}, 241306 (2013), arXiv:1302.4438.

\bibitem{Blennow:2013jba}
M.~Blennow, E.~Fernandez-Martinez, and B.~Zaldivar,
\newblock (2013), arXiv:1309.7348.

\bibitem{Hannestad:2004px}
S.~Hannestad,
\newblock Phys.Rev. {\bf D70}, 043506 (2004), arXiv:astro-ph/0403291.

\bibitem{Masso:2005ym}
E.~Masso and J.~Redondo,
\newblock JCAP {\bf 0509}, 015 (2005), arXiv:hep-ph/0504202.

\bibitem{Dienes:1996zr}
K.~R. Dienes, C.~F. Kolda, and J.~March-Russell,
\newblock Nucl. Phys. {\bf B492}, 104 (1997), arXiv:hep-ph/9610479.

\bibitem{Lukas:1999nh}
A.~Lukas and K.~Stelle,
\newblock JHEP {\bf 0001}, 010 (2000), arXiv:hep-th/9911156.

\bibitem{Abel:2003ue}
S.~A. Abel and B.~W. Schofield,
\newblock Nucl. Phys. {\bf B685}, 150 (2004), arXiv:hep-th/0311051.

\bibitem{Blumenhagen:2005ga}
R.~Blumenhagen, G.~Honecker, and T.~Weigand,
\newblock JHEP {\bf 0506}, 020 (2005), arXiv:hep-th/0504232.

\bibitem{Abel:2006qt}
S.~A. Abel, J.~Jaeckel, V.~V. Khoze, and A.~Ringwald,
\newblock Phys. Lett. {\bf B666}, 66 (2008), arXiv:hep-ph/0608248.

\bibitem{Abel:2008ai}
S.~A. Abel, M.~D. Goodsell, J.~Jaeckel, V.~V. Khoze, and A.~Ringwald,
\newblock JHEP {\bf 07}, 124 (2008), arXiv:0803.1449.

\bibitem{Goodsell:2009pi}
M.~Goodsell,
\newblock p. 165 (2009), arXiv:0912.4206.

\bibitem{Goodsell:2009xc}
M.~Goodsell, J.~Jaeckel, J.~Redondo, and A.~Ringwald,
\newblock JHEP {\bf 11}, 027 (2009), arXiv:0909.0515.

\bibitem{Goodsell:2010ie}
M.~Goodsell and A.~Ringwald,
\newblock Fortsch. Phys. {\bf 58}, 716 (2010), arXiv:1002.1840.

\bibitem{Heckman:2010fh}
J.~J. Heckman and C.~Vafa,
\newblock Phys.Rev. {\bf D83}, 026006 (2011), arXiv:1006.5459.

\bibitem{Bullimore:2010aj}
M.~Bullimore, J.~P. Conlon, and L.~T. Witkowski,
\newblock JHEP {\bf 1011}, 142 (2010), arXiv:1009.2380.

\bibitem{Cicoli:2011yh}
M.~Cicoli, M.~Goodsell, J.~Jaeckel, and A.~Ringwald,
\newblock JHEP {\bf 1107}, 114 (2011), arXiv:1103.3705.

\bibitem{Goodsell:2011wn}
M.~Goodsell, S.~Ramos-Sanchez, and A.~Ringwald,
\newblock JHEP {\bf 1201}, 021 (2012), arXiv:1110.6901.

\bibitem{Holdom:1985ag}
B.~Holdom,
\newblock Phys. Lett. {\bf B166}, 196 (1986).

\bibitem{Jaeckel:2013ija}
J.~Jaeckel,
\newblock Frascati Phys.Ser. {\bf 56}, 172 (2012), arXiv:1303.1821.

\bibitem{Arvanitaki:2009fg}
A.~Arvanitaki, S.~Dimopoulos, S.~Dubovsky, N.~Kaloper, and J.~March-Russell,
\newblock Phys. Rev. {\bf D81}, 123530 (2010), arXiv:0905.4720.

\bibitem{Acharya:2010zx}
B.~S. Acharya, K.~Bobkov, and P.~Kumar,
\newblock JHEP {\bf 1011}, 105 (2010), arXiv:1004.5138.

\bibitem{Cicoli:2012sz}
M.~Cicoli, M.~Goodsell, and A.~Ringwald,
\newblock JHEP {\bf 1210}, 146 (2012), arXiv:1206.0819.

\bibitem{Balasubramanian:2005zx}
V.~Balasubramanian, P.~Berglund, J.~P. Conlon, and F.~Quevedo,
\newblock JHEP {\bf 03}, 007 (2005), arXiv:hep-th/0502058.

\bibitem{Blumenhagen:2009gk}
R.~Blumenhagen, J.~Conlon, S.~Krippendorf, S.~Moster, and F.~Quevedo,
\newblock JHEP {\bf 0909}, 007 (2009), arXiv:0906.3297.

\bibitem{Cicoli:2011qg}
M.~Cicoli, C.~Mayrhofer, and R.~Valandro,
\newblock JHEP {\bf 1202}, 062 (2012), arXiv:1110.3333.

\bibitem{DeAngelis:2007dy}
A.~De~Angelis, O.~Mansutti, and M.~Roncadelli,
\newblock Phys. Rev. {\bf D76}, 121301 (2007), arXiv:0707.4312.

\bibitem{Meyer:2013pny}
M.~Meyer, D.~Horns, and M.~Raue,
\newblock Phys.Rev. {\bf D87}, 035027 (2013), arXiv:1302.1208.

\bibitem{Armengaud:2014gea}
E.~Armengaud {\em et~al.},
\newblock (2014), arXiv:1401.3233.

\bibitem{Bahre:2013ywa}
R.~B{\"a}hre {\em et~al.},
\newblock JINST {\bf 8}, T09001 (2013), arXiv:1302.5647.

\bibitem{Asztalos:2003px}
[ADMX Collaboration], S.~J. Asztalos {\em et~al.},
\newblock Phys. Rev. {\bf D69}, 011101 (2004), arXiv:astro-ph/0310042.

\bibitem{Sikivie:2013laa}
P.~Sikivie, N.~Sullivan, and D.~Tanner,
\newblock (2013), arXiv:1310.8545.

\bibitem{Budker:2013hfa}
D.~Budker, P.~W. Graham, M.~Ledbetter, S.~Rajendran, and A.~Sushkov,
\newblock (2013), arXiv:1306.6089.

\end{thebibliography}
\end{document}